\documentclass[]{PoS}

\usepackage{amsmath}

\def\Dslash{\not{\hbox{\kern-2.5pt $D$}}}

\def\mui{\mu_s}
\def\muf{\mu_f}
\def\muh{\mu_h}
\def\rgel{\mathsf{L}}
\newcommand{\vect}[1]{\mathbf{#1}}
\newcommand{\abs}[1]{\left\lvert #1\right\rvert}
\newcommand{\Lqcd}{\Lambda_{\textrm{QCD}}}
\newcommand{\e}{\mathrm{e}}
\newcommand{\as}{\alpha_s}

\title{Soft Collinear Effective Theory: An Overview}

\ShortTitle{SCET Overview}

\author{\speaker{Sean Fleming}\\
        University of Arizona\\
        E-mail: \email{fleming@physics.arizona.edu}}

\abstract{In this talk I give an overview of soft collinear effective theory (SCET), including a discussion of some recent advances. First, I briefly cover the foundation upon which SCET is built, namely QCD factorization, and review the theoretical framework of SCET. Next, I cover some recent calculations involving SCET. Since there are more interesting results than I can cover, I have picked four that that I find particularly exciting. The first two topics are dynamical threshold enhancements in Drell-Yan and  resummation of higgs production at the LHC. Both topics center on the resummation of large Sudakov logarithms, however, for each there is a twist. In the case of threshold enhancements in Drell-Yan, it is partonic, not hadronic logarithms that are summed. In the case of higgs production at the LHC, the twist is that the logarithms are of a time-like scale. The third topic highlighted is electroweak corrections at high energy. In particular the electroweak Sudakov form factor. Finally, I end by discussing generalized event shapes in $e^+e^-$ annihilation. }

\FullConference{International Workshop on Effective Field Theories: from the pion to the upsilon \\
		February 2-6 2009\\
		Valencia, Spain}

\begin{document}

\section{Introduction}
Effective field theories provide a simple and elegant method for calculating processes with several relevant 
energy scales~\cite{Weinberg:1978kz,Witten:kx,Georgi:qn,Harvey:ya,Manohar:1995xr,Kaplan:1995uv}.  
Part of the utility of effective theories is that 
they dramatically simplify the summation of logarithms of
ratios of mass scales, which would otherwise make perturbation theory
poorly behaved.  Furthermore, the systematic power counting in effective
theories, and the approximate symmetries of the effective field theory
can greatly reduce the complexity of calculations.

In this talk, I first review soft-collinear effective theory (SCET)~\cite{Bauer:2000ew,Bauer:2000yr,Bauer:2001ct,Bauer:2001yt}, 
which is an effective field theory describing the dynamics of highly energetic particles moving close to the light-cone 
interacting with a background field of soft quanta. Following this, I cover what are, in my opinion, some of the more interesting
recent applications of SCET.

The roots of SCET lie in the extensive foundations of QCD factorization theorems whose development goes back to the prehistory of QCD\footnote{For a detailed review see Ref.~\cite{Sterman:1995fz}.}. In essence, the goal of factorization is to systematically separate long and short distance dynamics in 
interactions, which allows for perturbative computations of the short-distance quantities, and a phenomenological 
determination of long-distance contributions. The key is the ability to identify sources of long distance 
or infrared (IR) behavior in perturbation theory. 

\begin{figure}
\begin{center}
\includegraphics[width=1.5in]{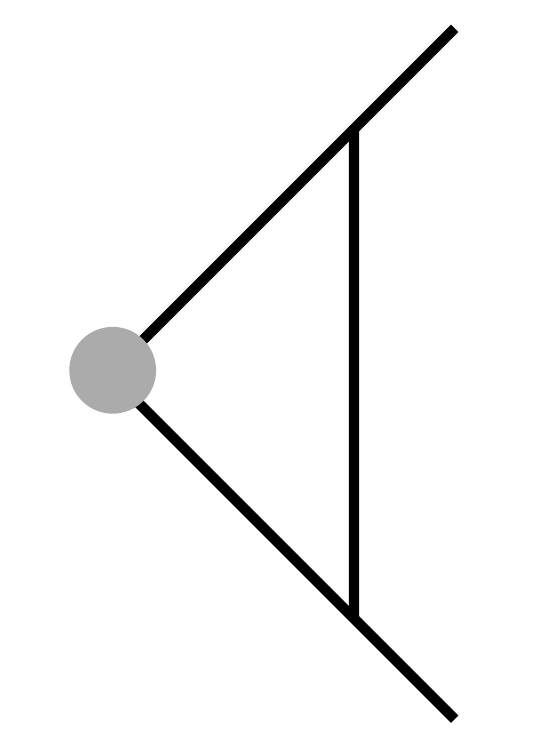}
\caption{Scalar vertex}
\label{scalarvertex}
\end{center}
\end{figure}

To understand better the nature of long distance behavior in perturbation theory let us consider a concrete example: the 
scalar vertex at one loop shown in Fig.~\ref{scalarvertex}.
This diagram is infrared divergent, and in dimensional regularization these IR divergences have the form
\begin{equation}
A_\textrm{\tiny sv}^\textrm{\tiny div} = \frac{1}{(-\epsilon)^2} - \frac{3}{2(-\epsilon)} \,.
\label{divsv}
\end{equation}
Notice that there are both single and double poles present. The question is: where do these poles come from? 
In the one loop diagram we are considering, there are two possible sources of IR divergences 
for massless particles in Minkowski space. 
The first source of IR divergence is when all components of the loop momentum go to zero: $k^\mu \to 0$. This is 
referred to as the soft region of the loop integral. The second source of IR divergences is when the loop momentum
is light-like. In particular if we define light-cone momenta $k^- = k^0-k^3$ and $k^+ = k^0+k^3$, where 
$k^2 = k^+ k^- - k_\perp^2$, then an IR divergence
can arise for 
\begin{equation}
k^- \sim \textrm{fixed} \qquad k^+ \to 0 \qquad k_\perp \to 0 \,.
\end{equation}
This is referred to as the collinear region of the loop momentum. Of course, we can also have a collinear region for $k^+$ 
fixed and $k^- \to 0$. Furthermore, there are overlapping regions where the loop momentum is both collinear and soft. 
This region gives rise to the double pole in Eq.~(\ref{divsv}).

The scaling of soft and collinear momentum can easily be parameterized by a single variable $\lambda$ which tends to 
zero. Two choices of such a parameterization are shown in Tab.~\ref{scaling}, where $Q$ is a large momentum: $Q \gg \Lqcd$.
\begin{table}[htdp]
\begin{center}
\begin{tabular}{ccc} 
& I & II \\
\hline \\
Soft & $k^\mu \sim \lambda^2 Q$ & $k^\mu \sim \lambda Q$ \\ 
 & &  \\
 & $k^-\sim Q$ &  \\
Collinear & $k^+ \sim \lambda^2 Q$ &  \\ 
& $k_\perp \sim \lambda Q$ & 
\end{tabular} 
    \caption{Two possible parameterizations of soft and collinear momentum}
\end{center}
\label{scaling}
\end{table}
Here we have chosen to fix the parameterization of the collinear momentum, and show two choices for the soft 
momentum. The soft momentum scaling in column I will henceforth be referred to as ultra-soft or usoft, while the scaling in column II will continued to be referred to as soft.

\begin{figure}
\begin{center}
\includegraphics[width=4in]{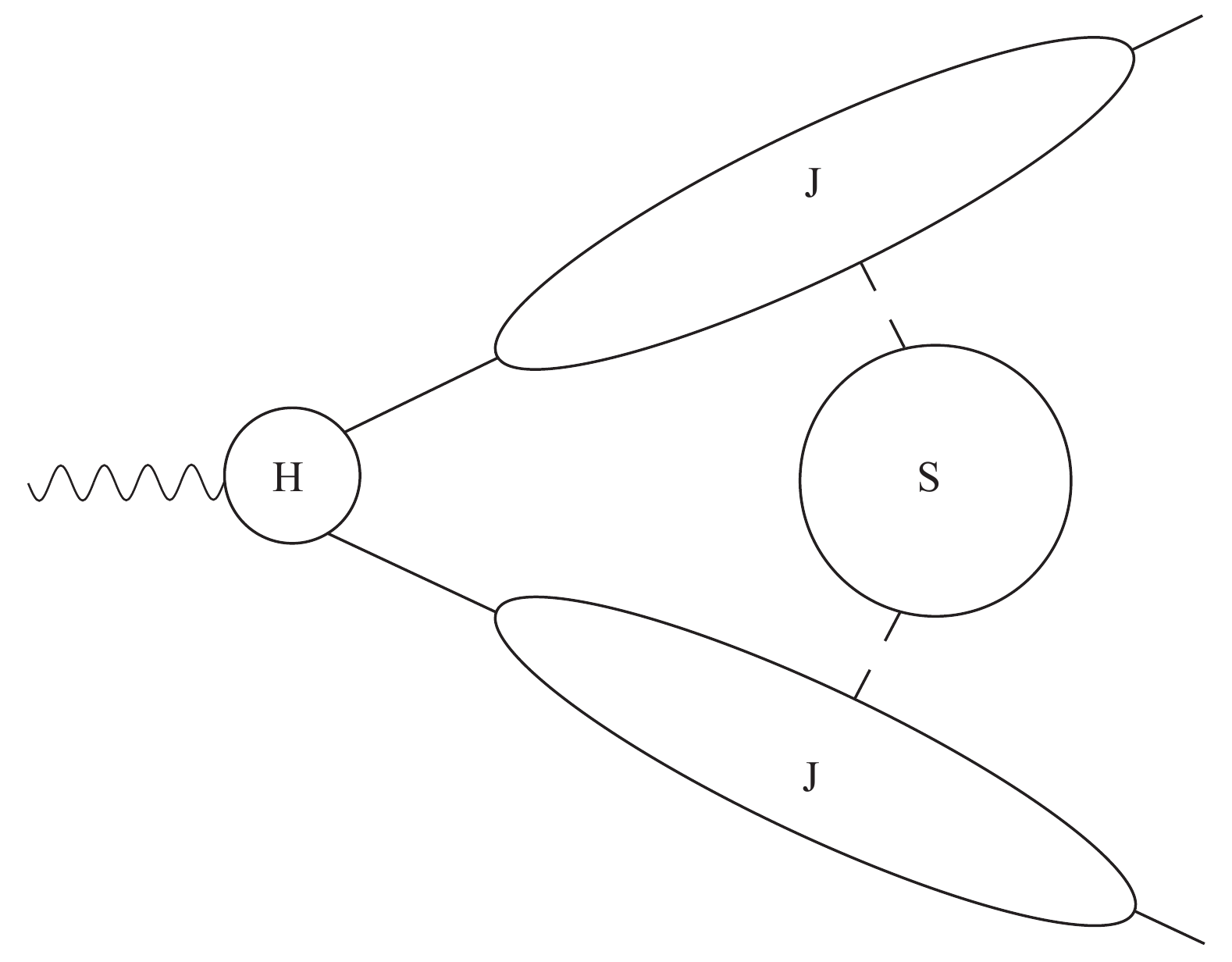}
\caption{Reduced diagram of the EM form factor, which shows the sources of IR sensitivity. }
\label{REMFF}
\end{center}
\end{figure}

The analysis of the one loop scalar vertex reveals the origin of the IR divergences arising in that calculation. 
There is no apriori reason, however, that the simple picture that emerges at one loop persists to all orders in perturbation theory.
As it turns out, the situation at all orders is exactly the same as at one loop: IR sensitive behavior comes from regions
of soft and collinear momentum~\cite{GS,Sterman:1978bi}. This result comes from an all orders analysis using the so-called Landau equations~\cite{ELOP,Landau:1959fi}, which 
reveals that, at any order a scattering process can be represented by a 
reduced diagram which describes the IR behavior of the interaction. For example, the reduced diagram of the 
electromagnetic (EM) form factor is shown in Fig.~\ref{REMFF}.
Here H stands for the hard contribution which only consists of momenta of order $Q$, J (or the jet) represents the contribution from collinear momenta, and S represents the contribution from soft momenta. There are two collinear directions (one along $\hat{z}$ and one along $-\hat{z}$), hence there are two collinear factors. The interpretation of the diagram is that hard interactions are short-distance and local. These hard interactions can produce particles with collinear momentum represented by the solid line, however they can not give rise to particles with soft momentum. Those particles with collinear momentum interact among themselves (any interactions with hard particles not shown are power suppressed by at least $1/Q$), and they can interact with particles of soft/usoft momenta, as indicated by the dashed lines. Of course soft/usoft particles can interact among themselves. Note, only soft/usoft momenta is communicated from the collinear jet in the $\hat{z}$ to the collinear jet in the $-\hat{z}$ direction.

\section{Soft Collinear Effective Theory}

This brings us to soft collinear effective theory. SCET is recasting perturbative factorization as an effective field theory. The motivation for this is two fold: first the EFT formalism allows for an extension of factorization to subleading order in $\lambda$ in a straightforward manner, and second naturally gives operator definitions, which hold non-perturbatively. I also, wish to point out that the formulation of SCET would not have been possible without serious advance in our understanding of EFTs, which surprisingly came out of studies of non-relativistic EFTs~\cite{Luke:1996hj,Grinstein:1997gv,Luke:1997ys,Luke:1999kz}.

Consider once again, the EM form factor. In the center-of-mass (COM) frame the light
particles in one of the jets move close to the light cone direction $n^\mu$ and their dynamics is
best described in terms of light cone coordinates $p = (p^+, p^-, p_\perp)$,
where $p^+ = n \cdot p$, $p^- = \bar{n} \cdot p$.  For large
energies the different light cone components are widely separated, with $p^-
\sim Q$ being large, while $p_\perp$ and $p^+$ are small. Taking the small
parameter to be $\lambda \sim p_\perp/p^-$ we have
\begin{equation} \label{collmom}
 p^\mu = \bar{n} \cdot p\, \frac{n^\mu}{2} + p_\perp^\mu + 
   n\cdot p\,\frac{\bar{n}^\mu}{2} = {\cal O}(\lambda^0) +{\cal O}(\lambda^1) +
   {\cal O}(\lambda^2) \,,
\end{equation}
where we have used $p^+p^- \sim p_\perp^2 \sim Q^2 \lambda^2$ for fluctuations near
the mass shell.  Thus the light-cone momentum components of collinear particles
scale like $k_c = Q(\lambda^2, 1,\lambda)$. The collinear quark can emit either 
a gluon collinear to the large momentum direction or a gluon with momentum 
scaling $k_{us} = Q(\lambda^2,\lambda^2,\lambda^2)$. For scales above the typical off-shellness of the collinear
degrees of freedom, $k_c^2 \sim (Q \lambda)^2$, both gluon modes
are required to correctly reproduce all the infrared physics of QCD~\cite{Bauer:2000ew}. Note, the scaling chosen above corresponds to
column I in Tab.~\ref{scaling}, and gives rise to a version of SCET often referred to as
SCET$_\textrm{\small I}$. The scaling in column II gives a different EFT referred to as 
SCET$_\textrm{\small II}$. From here-on out, when I do not specify the version of SCET
I mean SCET$_\textrm{\small I}$.
 
The SCET Lagrangian can be obtained at tree level by expanding
the full theory Lagrangian in powers 
of $\lambda$~\cite{Bauer:2000yr}, and 
the leading SCET Lagrangian for the collinear quark sector is 
\begin{equation}\label{scetlag}
{\cal  L}^{(0)}_{\xi \xi} = \bar{\xi}_{n,p'} \bigg\{ i n\cdot D + i {D\!\!\!\!\slash}^\perp_c
\frac{1}{i\bar{n} \cdot D_c} i   {D\!\!\!\!\slash}^\perp_c \bigg\} \frac{{\bar n\!\!\!\slash}}{2}
\xi_{n,p}\,,
\end{equation}
where $i n\cdot D = i n \cdot \partial + g n\cdot A_{n,q} + g n\cdot A_{us}$,  
$i\bar{n} \cdot D_c = \bar{{\cal P}} + g \bar{n}\cdot A_{n,q}$, and 
$i D^{\perp \mu}_c = {\cal P}^{\perp\mu} + g A^{\perp\mu}_{n,q}$. 
Here we introduce a projection
operator ${\cal P}$ which only acts on the large label of the collinear 
fields~\cite{Bauer:2001ct}. For any function $f$
\begin{eqnarray}
&& f(\bar{{\cal P}}) \phi^\dagger_{q_1} \cdots \phi^\dagger_{q_m} 
\phi_{p_1} \cdots \phi_{p_n} 
\nonumber \\
&& \hspace{3ex} f(\bar{n} \cdot p_1 + \dots + \bar{n} \cdot p_n - 
\bar{n} \cdot q_1 - \dots \bar{n} - \cdot q_m)
\phi^\dagger_{q_1} \cdots \phi^\dagger_{q_m} 
\phi_{p_1} \cdots \phi_{p_n} \,,
\end{eqnarray}
where $\bar{{\cal P}} \equiv \bar{n} \cdot {\cal P}$. 

An important aspect of effective field
theories is the approximate symmetries that are manifest in the leading order
Lagrangian. The SCET Lagrangian presented above has a global helicity spin 
symmetry. In addition, 
SCET has a powerful set of gauge symmetries~\cite{Bauer:2001yt}. 
Specifically the collinear and usoft fields each have their own gauge transformation that leave the 
Lagrangian invariant. Collinear gauge transformations are the subset of QCD
gauge transformations where $\partial^\mu U(x) \sim Q (\lambda^2, 1 , \lambda)$,
and usoft gauge transformations are those where $\partial^\mu V(x) \sim  Q \lambda^2$. 
The invariance under each of these transformations is a manifestation of 
scales of order $Q$ or greater having been removed from the theory, since 
any gauge transformation that would change a usoft gluon into a collinear gluon 
would imply a boost of order $Q$. 

The SCET Lagrangian describes the interactions of collinear and soft quanta, while all of the hard physics, involving momenta of order $Q$ has been "integrated out". This is the case in operators as well. Operators in QCD are matched onto products of SCET operators and Wilson coefficients. Since the SCET operators correctly describe the IR physics of the QCD operators, but get the short distance physics wrong, the Wilson coefficients act as correction factors so that the right short-distance behavior of the QCD operators is reproduced. In the language of QCD factorization, this is just the factorization of hard from collinear and soft degrees of freedom. 

Factoring the usoft degrees of freedom from the collinear degrees of freedom requires something else. Note, the Lagrangian in Eq.~(\ref{scetlag}) still contains the term $g n\cdot A_{us}$ that couples collinear quarks to usoft gluons. This term can be removed by redefining the collinear fields~\cite{Bauer:2001yt}:
\begin{eqnarray}
\xi_{n,p}(x) & \to & Y_n(x) \xi^{(0)}_{n,p}(x)
\nonumber \\
A_{n,q}(x) & \to & Y_n(x)A^{(0)}_{n,q}(x)Y^\dagger_n(x) \,,
\end{eqnarray}
with
\begin{equation}
Y_n(x) = \textrm{P} \, \textrm{exp}\bigg( i g \int^{\infty}_0 ds \, n\cdot A_{us}(ns + x) \bigg) \,,
\end{equation}
where $Y_n$ is a path ordered exponential commonly referred to as a Wilson line. As a consequence of the field redefinition the collinear sector of the Lagrangian decouples from the usoft sector, and at leading order SCET becomes a direct product theory of collinear and usoft with no interactions between them. This leads to the factorization of usoft and collinear in operators, which along with the Wilson coefficient reproduces QCD factorization results.

\section{Recent Results}

Next, I discuss four recent calculation using SCET that I find particularly intriguing. The first two results use methods developed for SCET to understand the resummation of large logarithms. This by-itself is not novel, but the nature of the logarithms being summed is a bit different than usual. In the case of dynamical threshold enhancement in Drell-Yan, the important logarithms occur at partonic threshold, as opposed to the hadronic threshold. This makes the justification of a resummation of great interest. In resummation for Higgs production at hadronic colliders, the logarithms in question contain a time-like scale. This time-like scale leads to a theoretically interesting issue, which is resolved. The second set of results I will discuss are on subjects that will be covered in other talks at this conference. Thus, I will only briefly cover the highlights of these calculations and refer the reader to the proceedings of the presenters covering each topic. The first of these two topics concerns the proper treatment of electroweak Sudakov logarithms in SCET. The second of the two topics concerns calculations of generalized event shapes in SCET.

\subsection{Dynamical Threshold Enhancement in Drell-Yan}

The Drell-Yan  process \cite{Drell:1970wh}, which is the production of a lepton pair in hadron-hadron collisions, has played an important role in establishing the parton picture underlying the strong interactions. Consequently, a lot of effort has been put in obtaining accurate theoretical predictions for Drell-Yan in perturbative QCD. Fixed order calculations have been carried out up to an impressive NNLO level~\cite{Altarelli:1979ub,Hamberg:1990np,Harlander:2002wh,Anastasiou:2003yy,Anastasiou:2003ds,Melnikov:2006di,Melnikov:2006kv}, and additional improvements in the form of all order perturbative resummations have been made as well. The need for these resummations arises when the invariant mass $M$ of the lepton pair approaches the COM energy of the collision. In this regime there is limited phase space available for the emission of QCD radiation, and large Sudakov logarithms involving the ratio of the scale $M$ and a soft scale $\mui \ll M$ remain. These ``threshold logarithms" threaten the convergence of the perturbative expansion and need to be resummed to all orders. Resummation of threshold logarithms has been carried out to next-to-next-to-next-to-leading logarithmic (N$^3$LL) order~\cite{Sterman:1986aj,Catani:1989ne,Magnea:1990qg,Korchemsky:1993uz,Sterman:2000pt,Mukherjee:2006uu,Bolzoni:2006ky,Ravindran:2006bu,Ravindran:2007sv}. 

Until the recent work of Ref.~\cite{Becher:2007ty}, however, there has not been a formal justification for resummation. The reason is, the PDFs are strongly suppressed in the endpoint region $x\to 1$, so the cross section $d\sigma/dM^2$ is a  steeply falling function as $M$ approaches the kinematic endpoint $\sqrt{s}$, and in a typical experiment it is not be possible to observe Drell-Yan pairs with masses exceeding about one half of the COM energy. In practice, one is therefore never in a region where the ratio $\tau=M^2/s$ approaches 1. Since threshold resummation deals with logarithms of the form $\ln(1-\tau)$, it is then not obvious why such terms should be treated on different footing than other higher-order terms.  A heuristic argument why threshold resummation effects could be important even if $\tau\ll 1$ is given in Refs.~\cite{Appell:1988ie,Catani:1998tm}. The idea is that the sharp fall-off of the parton luminosity at large $x$ dynamically enhances the contribution of the {\em partonic\/} threshold region $z=M^2/\hat s\to 1$, i.e., the region where the COM energy $\sqrt{\hat s}$ of the initial-state partons is just sufficiently large to produce the Drell-Yan pair. It is argued that it could then be important to resum logarithms of the form $\ln(1-z)$ in the hard partonic cross section. However, since $(1-z)$ is not related to a small ratio of external physical scales, it is not obvious how to give a formal justification of this argument. 

This last question was studied quantitatively in Ref.~\cite{Becher:2007ty} in the context of SCET. The authors found that in the true endpoint region $\tau\to 1$, the effective soft scale $\mui$ is an order of magnitude smaller than the naive guess $M(1-\tau)$. For PDFs behaving like $f_{i/N}(x)\sim(1-x)^{b_i}$ near $x\to 1$, they find $\mui\approx\lambda^{-1} M(1-\tau)$ with $\lambda\approx 2+b_q+b_{\bar q}=O(10)$. This result provides a formal justification to the argument of a dynamical enhancement of the partonic threshold region due to the fall-off of parton densities. They also found that the dynamical enhancement of the threshold contributions remains effective down to moderate values $\tau\approx 0.2$, while at very small $\tau$ values the parameter $\lambda$ decreases to about 2. This reflects the fact that for small $x$ values the fall-off of the PDFs is much weaker than for large $x$. In fact even far away from the true threshold the Drell-Yan cross section receives its dominant contributions from those terms in the hard partonic cross section that are leading in the limit $z\to 1$. Another result is that with the appropriate choice of the effective soft scale $\mui$, the convergence of the perturbative expansion is greatly improved by resummation. The authors find, however, that for small enough $M$ the terms beyond $O(\alpha_s^2)$ in the resummed expression for the cross section are numerically unimportant, but for larger masses the effects can be significant. For instance, the experiment E866/NuSea has reported data up to $M=16.85$\,GeV (corresponding to $\tau\approx 0.19$) \cite{Webb:2003ps}, and at $M=16$\,GeV it is found that resummation effects enhance the fixed-order predictions for the cross section by about 25\% at NLO, and 7\% at NNLO. 

\begin{figure}[t!!]
\centerline{
\includegraphics[width=6in]{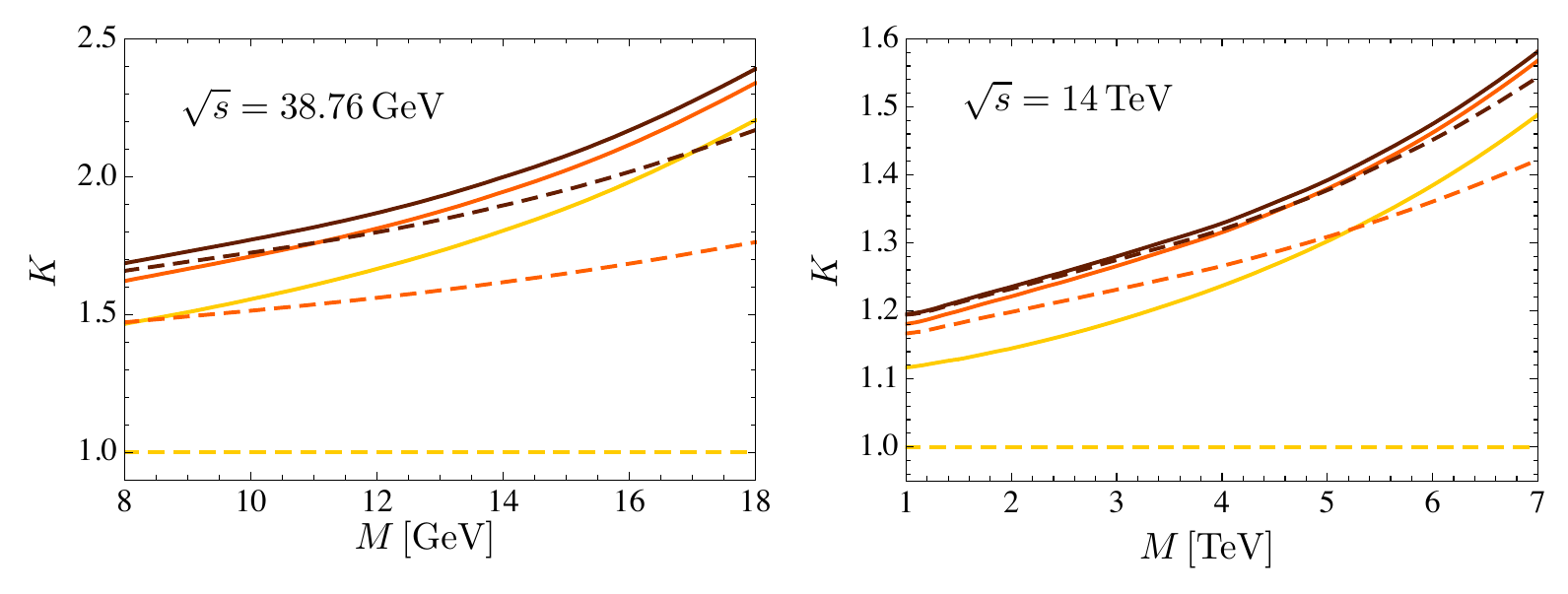}
}
\caption{\label{fig:DY} Convergence of the DY $K$-factor at 
threshold: dashed lines refer to the fixed-order
calculation (from bottom to top: LO, NLO, NNLO); solid lines
to the corresponding resummed result.}
\end{figure}

To be specific, consider the DY cross section in the threshold region, $z= M^2/\hat s \to 1$. To leading order the cross section has a factored form:
\begin{equation}
\frac{d\sigma^{\rm thr.}}{dM^2} 
 \propto \, 
 \sum_q e_q^2 \int \frac{dx_1}{x_1} \, \frac{dx_2}{x_2} 
  \, \theta[\hat s - M^2]\,
  { C(z,M;\mu_f)} 
  {\left[ f_{q/N_1}(x_1;\mu_f) \, f_{\bar q/N_2}(x_2;\mu_f) + (q \leftrightarrow \bar q) \right]}\,.
\end{equation}
where 
$x_{1,2}$ are the parton momentum fractions, and
$\hat s = x_1 x_2 s$.
For small values of $(1-z)$, the short distance coefficient may be factored further \cite{Becher:2007ty}
\begin{equation}
  {C(z,M;\mu_f) = {} } {H(M,\mu_f)} \, {S(\sqrt{\hat s}\,(1-z);\mu_f)} \,.
\end{equation}
This separates the effects associated with
the hard scale, $M^2 \sim \hat s$, set by the partonic sub-process, from 
the collinear scale, $(1-z) M^2$, related to the virtuality of the colliding partons,
and the soft scale, $(1-z)^2 M^2$, related to the invariant mass of the hadronic remnants.
In the endpoint region it is sufficient to assume a simple parametrization for the quark PDFs
at large momentum fraction,
$
 f_{q/N}(x)\big|_{x \to 1} = N_q \, (1-x)^{b_q}
$.
It is shown in Ref.~\cite{Becher:2007ty}, that DY-production at threshold is dominated by $d$\/-quarks 
(which have the largest value of $b_q$), and that the 
resummed $K$-factor can be written in analytic form. From this analysis
one deduces the appearance of an effective soft scale
\cite{Becher:2007ty},
\begin{equation}
\mu_s \sim \frac{M \, (1-\tau)}{2 + b_d + b_{\bar d}} 
  \approx  \frac{M \, (1-\tau)}{13} \,.
\end{equation}
As shown in Fig.~\ref{fig:DY} the perturbative convergence of the $K$-factor 
is significantly improved compared to the fixed order results.

\begin{figure}
\begin{center}
\includegraphics[width=5.9in]{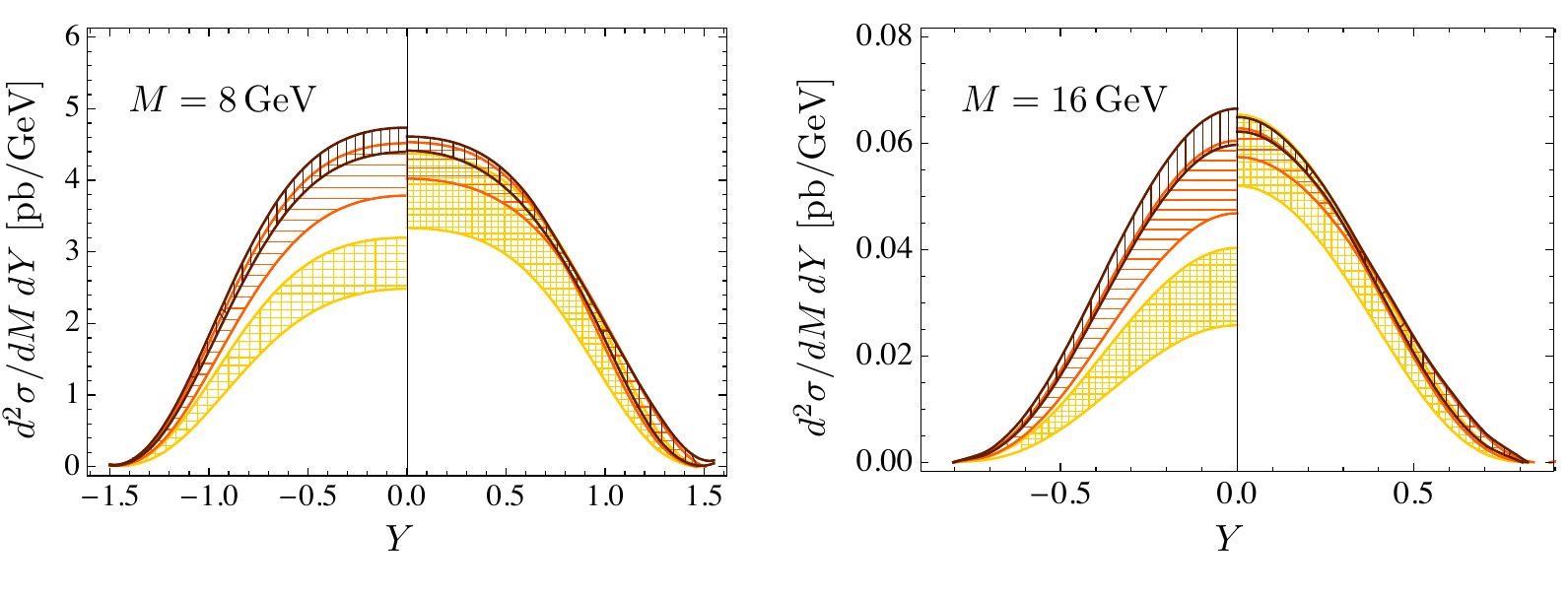}
\caption{The Drell-Yan resummed rapidity distribution (right-hand side of each graph) versus the fixed order result (left-hand side of each graph) at $\sqrt{s}=38.76$\,GeV. The left graph is for $M=8$\,GeV, and the right graph is for $M=16$\,GeV. The light bands are LO, medium bands are NLO, and dark bands are NNLO.}
\label{DYRSRD}
\end{center}
\end{figure}

In my opinion, the most interesting result of Ref.~\cite{Becher:2007ty} is the rapidity distribution at $\sqrt{s}=38.76$\,GeV. Note, for this energy and $M=8$\,GeV,  $\tau\approx 0.04$ is very small. The predictions were made by
combining the resummed result for the cross section with the power-suppressed terms calculated in fixed-order perturbation theory:
\begin{equation}\label{match}
   \frac{d\sigma^{\rm combined}}{dM^2 dY} 
   = \left. \frac{d\sigma^{\rm thresh}}{dM^2 dY}
    \right|_{\muh,\mui,\muf}
   + \left( 
    \left. \frac{d\sigma^{\rm fixed\,order}}{dM^2 dY}\right|_{\muf} 
    - \left. \frac{d\sigma^{\rm thresh}}{dM^2 dY} 
    \right|_{\muh=\mui=\muf} \right) .
\end{equation}
In Fig.~\ref{DYRSRD} the RG-improved results are compared with the fixed-order results, with varying scales. The bands reflect the variations about the default value.
It is clear that the resummation significantly improves the convergence of the perturbative expansion. Moreover, in the resummed case the combined uncertainty at NLO and NNLO is much smaller than in the fixed-order case. 

\subsection{Resummation for Higgs Production at Hadronic Colliders}

Another interesting application of SCET methods is the work of Refs.~\cite{Ahrens:2008qu,Ahrens:2008nc} on resummation for Higgs production at hadronic colliders. What I think is fun about this calculation is that large logarithms involving a time-like, rather than space-like scale are resummed. 

The Higgs-boson production cross section at hadron colliders such as the Tevatron or LHC is dominated by the gluon-gluon fusion process $gg\to H$ via a top-quark loop, which for light Higgs is well approximated by the effective local interaction \cite{Inami:1982xt}
\begin{equation}\label{Heff}
   {\cal L}_{\rm eff} = C_t(m_t^2,\mu^2)\,\frac{H}{v}\,
   G_{\mu\nu,a}\,G_a^{\mu\nu} \,,
\end{equation} 
where $v\approx 246$\,GeV is the Higgs vacuum expectation value, and the short-distance coefficient $C_t=\alpha_s/(12\pi)+{\cal O}(\alpha_s^2)$ is known to NNLO  \cite{Chetyrkin:1997iv} and has a well behaved perturbative expansion for $\mu\sim m_H$. The production cross section is related to the discontinuity of the product of two such effective vertices. It can be written as the convolution of a hard-scattering kernel with parton distribution functions. Large corrections  due to virtual corrections to the effective $ggH$ interaction (\ref{Heff}) arise from quantum corrections characterized by the scale $\mu\sim m_H$. These effects are described by a universal factor and affect differential distributions in same way as the total cross section. They can be factorized into a hard function $H(m_H^2,\mu^2)$, which is the square of the on-shell gluon form factor evaluated at time-like momentum transfer $q^2=m_H^2$, and with IR divergences subtracted using the $\overline{\rm MS}$ scheme \cite{Becher:2007ty,Becher:2006nr,Idilbi:2006dg}. On a  technical level, the hard function appears as a Wilson coefficient in the matching of the two-gluon operator in (\ref{Heff}) onto an operator in SCET~\cite{Bauer:2001yt,Bauer:2002nz}, in which all hard modes have been integrated out. This matching takes the form
\begin{equation}\label{CSdef}
   G_{\mu\nu,a}\,G_a^{\mu\nu}
   \to C_S(Q^2,\mu^2)\,Q^2\,g_{\mu\nu}\,
   {\cal A}_{n\perp}^{\mu,a}\,
   {\cal A}_{\bar n\perp}^{\nu,a} \,,
\end{equation}
where $Q^2=-q^2$ is (minus) the square of the total momentum carried by the operator. The fields ${\cal A}_{n\perp}^{\mu,a}$ and ${\cal A}_{\bar n\perp}^{\nu,a}$ are effective, gauge-invariant gluon fields in SCET \cite{Hill:2002vw}. They describe  gluons propagating along the two light-like directions $n,\bar n$ defined by the colliding hadrons. 

The two-loop expression for the Wilson coefficient $C_S$ can be extracted from the results of \cite{Harlander:2000mg}, and has the form
\begin{equation}\label{CSexp}
   C_S(Q^2,\mu^2) = 1 + \sum_{n=1}^\infty\,c_n(L)
   \left( \frac{\alpha_s(\mu^2)}{4\pi} \right)^n \! ,
\end{equation}
where $L=\ln(Q^2/\mu^2)$. The one-loop coefficient is
\begin{equation}\label{c1c2}
   c_1(L) = C_A \left( -L^2 + \frac{\pi^2}{6} \right) ,
\end{equation}
and the result for the two-loop coefficient can be found in \cite{Ahrens:2008qu,Idilbi:2006dg}. The hard function is given by the absolute square of the Wilson coefficient at time-like momentum transfer, 
\begin{equation}\label{Hdef}
   H(m_H^2,\mu^2) = \left| C_S(-m_H^2-i\epsilon,\mu^2) \right|^2 .
\end{equation}

The Wilson coefficient at space-like momentum transfer has a well behaved expansion in powers of the coupling constant, if the renormalization scale is taken to be of order the natural scale, $\mu^2\sim Q^2$. For instance, with $N_c=3$ colors and $n_f=5$ light quark flavors:
\begin{equation}\label{CSeucl}
   C_S(Q^2,Q^2) = 1 + 0.393\,\alpha_s(Q^2) 
   - 0.152\,\alpha_s^2(Q^2) + \dots \,.
\end{equation} 
The nature of the expansion changes drastically when the same coefficient is evaluated at time-like momentum transfer $Q^2=-q^2-i\epsilon$:
\begin{eqnarray}\label{CSbad}
   C_S(-q^2,q^2) 
   &=& 1 + 2.75\,\alpha_s(q^2) + (4.84+2.07i)\,\alpha_s^2(q^2) 
    \nonumber\\
   &&\mbox{}+ \dots \,.
\end{eqnarray} 
The expansion coefficients are more than an order of magnitude larger than in the space-like region. The origin of this effect is that the Sudakov (double) logarithms contained in the coefficients $c_n(L)$ in (\ref{CSexp}) give rise to $\pi^2$ terms when we analytically continue $L\to\ln(q^2/\mu^2)-i\pi$. 

The large expansion coefficients in the perturbative series for the Wilson coefficient in the time-like region can be avoided if we evaluate this coefficient at a {\em time-like\/} renormalization point, in which case 
\begin{equation}\label{Ctimelike}
   C_S(-q^2,-\mu^2) = 1 + \sum_{n=1}^\infty\,c_n(L)
   \left( \frac{\alpha_s(-\mu^2)}{4\pi} \right)^n
\end{equation}
with $L=\ln(q^2/\mu^2)$ and the {\em same\/} expansion coefficients as in (\ref{CSexp}). One obtains
\begin{equation}\label{CSgood}
   C_S(-q^2,-q^2) = 1 + 0.393\,\alpha_s(-q^2) 
   - 0.152\,\alpha_s^2(-q^2) + \dots
\end{equation} 
instead of (\ref{CSbad}), which indeed exhibits a vastly better behavior. 

\begin{figure}
\includegraphics[width=0.45\columnwidth]{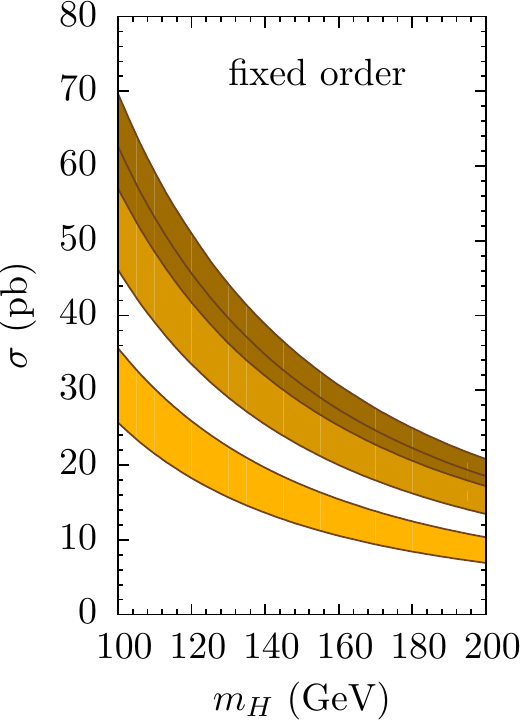} \quad
\includegraphics[width=0.45\columnwidth]{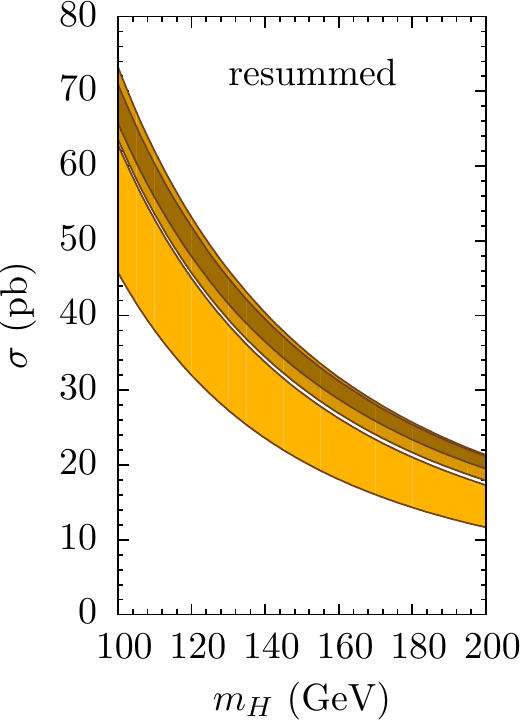}
\caption{\label{fig:sigma}
LO (light), NLO (medium), and NNLO (dark) predictions for the Higgs-production cross section at the LHC in fixed-order perturbation theory (left) and after resummation of the $\pi^2$-enhanced terms (right).}
\vspace{-0.4cm}
\end{figure}

In the expressions above, the running coupling is evaluated at time-like momentum transfer $-\mu^2-i\epsilon$. Since the function $\alpha_s(\mu^2)$ in perturbation theory is analytic in the complex $\mu^2$ plane (aside from a cut on the negative real axis and a  pole at $\mu^2=\Lambda_{\overline{\rm MS}}^2$) a running coupling at time-like argument in terms of that at space-like momentum transfer can be defined. At NLO Refs.~\cite{Ahrens:2008qu,Ahrens:2008nc} obtain
\begin{equation}\label{asNLO}
   \frac{\alpha_s(\mu^2)}{\alpha_s(-\mu^2)}
   = 1 - ia(\mu^2) 
   + \frac{\beta_1}{\beta_0}\,\frac{\alpha_s(\mu^2)}{4\pi}\, 
   \ln\left[ 1 - ia(\mu^2) \right] + {\cal O}(\alpha_s^2) \,,
\end{equation}
where $a(\mu^2)=\beta_0\alpha_s(\mu^2)/4$. The above equation is the key to resumming the logarithms of the time-like scale appearing in Higgs Production at Hadronic Colliders. This resummation is carried out in Refs.~\cite{Ahrens:2008qu,Ahrens:2008nc}, and the authors obtain improved results for the hard function in the formula for the Higgs-boson production cross section. Setting $\mu=m_H=120$\,GeV, they find
\begin{equation}
   H(m_H^2,m_H^2) = \{ 1.756_{\rm \,(LO)}, 1.907_{\rm \,(NLO)},
   1.906_{\rm \,(NNLO)} \} \,.
\end{equation}
This should be compared with the poorly converging series $H=\{1,1.623,1.844\}$ obtained using fixed-order perturbation theory. Fig.~\ref{fig:sigma} illustrates the impact of the resummation of the $\pi^2$-enhanced terms on the cross-section predictions for Higgs-boson production at the LHC. The bands in each plot show results obtained at LO, NLO, and NNLO using MRST2004 parton distributions \cite{Martin:2004ir}. Their width reflects the scale variation obtained by varying the factorization and renormalization scales between $m_H/2$ and $2m_H$ (setting $\mu_r=\mu_f$). The convergence of the expansion and the residual scale dependence at NLO and NNLO are greatly improved by the resummation. The new LO and NLO bands almost coincide with the NLO and NNLO bands in fixed-order perturbation theory, and the new NNLO band is now fully contained inside the NLO band. 

\subsection{Electroweak Corrections at High Energy}

This topic is the subject of the talk by A. Fuhrer, and I refer the interested reader to 
his conference proceedings for more details~\cite{Chiu:2009yz}. Here I will only review
the highlights of the work.

The Large Hadron Collider will be able to measure collisions with a partonic
COM energy of several TeV, which is more than an order of magnitude
larger than the masses of the electroweak gauge bosons. As a consequence, large 
Sudakov logarithms of the ratio of the two scales could ruin perturbative calculations.
Specifically, Electroweak Sudakov corrections will be roughly of the size $\alpha \log^2
(\hat{s}/M^2)_{W,Z}/(4 \pi \sin^2 \theta_W) \sim 0.15$ at
$\sqrt{\hat{s}}=4$~TeV, and these Sudakov corrections might need to 
be summed to all orders. The summation of electroweak Sudakov logarithms using effective field
theory methods has been discussed in detail in in Refs.~\cite{Chiu:2008vv,Chiu:2007dg,Chiu:2007yn} . 

The perturbation series for the logarithm of the Euclidean form-factor $F_E(Q^2)$ takes a simple form
\begin{eqnarray}
\log F_E &=&  \alpha\left(\tilde k_{12}\rgel^2+\tilde k_{11}\rgel +
\tilde k_{10}\right) +\alpha^2\left(\tilde k_{23}
\rgel^3+ \tilde k_{22}\rgel^2+ \tilde k_{21}\rgel+\tilde
k_{20}\right) +\ldots \,,
\label{2}
\end{eqnarray}
with $\rgel = \log(Q^2/M^2)$, the $\alpha^n$ term having powers of $\rgel$ up to $\rgel^{n+1}$, and the expansion begins at order $\alpha$. 
The right-hand-side of Eq.~(\ref{2}) can be written in terms of the LL series $\rgel f_0(\alpha \rgel)=\tilde k_{12} \alpha \rgel^2+
\tilde k_{23} \alpha^2 \rgel^3 + \ldots$, the NLL series $f_1(\alpha \rgel)=\tilde k_{11} \alpha \rgel +
\tilde k_{22} \alpha^2 \rgel^2 + \ldots$, the NNLL series $\alpha f_2(\alpha \rgel)=\tilde k_{10} \alpha  +
\tilde k_{21} \alpha^2 \rgel + \ldots$ etc.\ as
\begin{eqnarray}
\log F_E &=&\rgel f_{0}(\alpha \rgel)+f_{1}(\alpha \rgel)+\alpha f_{2}(\alpha \rgel)
+ \ldots\, .
\label{3}
\end{eqnarray}
$f_0$ and $f_1$ begin at order $\alpha$, and the remaining $f_n$ begin at order one.
Since for electroweak corrections at the TeV scale $\alpha\rgel^2$ is
quite sizeable, the $\textrm{LL}$ series can be summed up to all orders
 with the general result \cite{Chiu:2007yn}
\begin{equation}
 \ln F_E(Q) = C(Q) + \int_Q^M \frac{d\mu}{\mu} \left(
 \Gamma_{\rm cusp} L_Q + \gamma\right)
+ D(M) + \int_M^\mu \frac{d\mu}{\mu} \left(
 \tilde \Gamma_{\rm cusp} L_Q + \tilde \gamma \right) \,.
\end{equation}
Here $C(Q)$ is a matching coefficient at the high scale $Q$,
whose leading terms has the structure
\begin{equation}
 C(\mu)= \sum_{i=1}^3 \frac{\alpha_i(\mu)C_F^i}{4\pi} 
 \left[-L_Q^2 + \# L_Q + \# \right] + {\cal O}(\alpha_i^2) \,
\end{equation}
where $L_Q = \ln \frac{Q^2}{\mu^2}$, and 
$i=1..3$ refers to the three SM gauge group factors with
$C_F^i$ being the corresponding Casimirs. 
The numerical coefficients ($\#$) depend on the spin
of the two particles. Notice that $C(Q)$ does not depend on the
gauge-boson masses.
The RG-running between the high-energy scale $Q$ and the EW
gauge-boson mass scale $M \sim M_{W,Z}$ is controlled by the anomalous
dimension, which has a universal part, the cusp anomalous dimension
related to the Sudakov double logarithms, 
$ \Gamma_{\rm cusp}   = 4 \, \sum_{i=1}^3 \frac{\alpha_i C_F^i}{4\pi}  + {\cal O}(\alpha_i^2)$,
and a conventional part $\gamma$.

Similarly, $D(M)$ is the matching coefficient arising from integrating
out the massive gauge bosons in the SM, where the effective-theory
construction automatically takes care of the correct incorporation of
gauge-boson mixing,
 \begin{eqnarray} \label{IntMatch}
D(\mu) & = &  { \frac{\alpha_{\rm em}}{4\pi} 
 \frac{( T_3 - \sin^2\theta_W \, Q_{\rm em} )^2}{\sin^2\theta_W \, \cos^2\theta_W}}
\times 
 \left[ - L_{M_Z}^2 + 2 L_{M_Z} { L_Q} - \frac{5\pi^2}{6} \, + \, 
      \# L_{M_Z} + \# \right]
\cr 
   && \quad {} + {\frac{\alpha_{\rm em}}{4\pi} 
 \frac{T^2 - (T_3)^2}{\sin^2\theta_W}}
 \quad \times  \left[ - L_{M_W}^2 + 2 L_{M_W} {L_Q} - \frac{5\pi^2}{6} \, + \, 
      \# L_{M_W} + \# \right] + \ldots
\end{eqnarray}
A subtle point is the (single-logarithmic) 
dependence of the low-energy matching
coefficient on the high-energy scale via $L_Q$, which can be traced back to the
appearance of end-point singularities in individual diagrams \cite{Chiu:2007yn}. The appearance of the large logarithm $L_Q$ in the matching coefficient $D(\mu)$ would normally be of great concern, since this term is enhanced, and there is, in general no guarantee that terms of the form $(\alpha L_Q)^n$ would not arise at higher orders. However, it is proven in Ref.~\cite{Chiu:2007dg} that only a single power of $L_Q$ can arise, so higher order terms will indeed be suppressed by powers of $\alpha$ relative to the result in Eq.~(\ref{IntMatch}).
Finally, the RG-running in the SCET below the scale $M$ (via $\tilde \Gamma_{\rm cusp}$ 
and $\tilde\gamma$) is obtained by
replacing $\sum \alpha_i C_F^i \to \alpha_s C_F^{(3)} + \alpha_{\rm em} \, Q^2_{\rm em}$
(for QCD $\otimes$ QED).

\subsection{Event Shapes in $e^+ e^-$ annihilation}

This subsection gives the highlights of the talk by C. Lee~\cite{Lee:2009cw} on generalized event shapes, and I refer the interested reader to Lee's proceeding for more details. Event shapes yield simple information about the geometry of a hadronic final state in $e^+ e^-$ annihilation and can be used to probe the strong interactions at various energy scales~\cite{Dasgupta:2003iq}. Two-jet event shapes $e$ are designed so that they take a numerical value, usually between 0 and 1, so that one of the kinematic endpoints (usually $e=0$) corresponds to events with two perfectly-collimated back-to-back jets in the final state. Event shape distributions depend on the hard-scattering cross-section at the large center-of-mass energy $Q$, on the perturbative branching and showering of the hard partons into jets at intermediate scales, and on the soft color exchange between jets and hadronization at a soft scale $\Lqcd$. Event shapes are thus useful probes of both perturbative and nonperturbative effects in QCD, allowing, for instance, extraction of the strong coupling $\as$ and nonperturbative shape function parameters~\cite{Becher:2008cf}  (also see talks by I. Stewart and V. Mateu).

The most familiar  event shape is thrust, $T = \frac{1}{Q}\max_{\vect{t}}\sum_{i\in X} \abs{\vect{t}\cdot\vect{p}_i}$, 
where $Q$ is the $e^+ e^-$ center-of-mass energy, and  $\vect{t}$, the \emph{thrust axis}, is the unit three-vector which maximizes the sum of projections of final-state particles' three-momenta $\vect{p}_i$ onto this axis. Once the thrust axis is determined, many other event shapes can be defined, such as the jet broadening, $B = \frac{1}{Q}\sum_{i\in X} \abs{\vect{t}\times \vect{p}_i}$. 
A generalization of thrust and jet broadening is the class of \emph{angularities} \cite{Berger:2003iw},
\begin{equation}
\tau_a(X) = \frac{1}{Q}\sum_{i\in X}E_i \sin^a \theta_i (1-\cos\theta_i)^{1-a} = \frac{1}{Q}\sum_{i\in X} \abs{\vect{p}_i^T}\e^{-\abs{\eta_i}(1-a)}\,,
\end{equation}
where in the first form, $E_i$ is the energy of final-state particle $i$ and $\theta_i$ is its angle with respect to the thrust axis. In the second form, $\vect{p}_i^T$ is the $i$th particle's transverse momentum  and  $\eta_i$ its rapidity with respect to $\vect{t}$.  The parameter $a$ can be any real number, $-\infty<a<2$ for $\tau_a$ to be an infrared-safe observable. Two special cases are $a=0$ and $a=1$, which correspond to the thrust and jet broadening, $\tau_0 = 1 - T$ and $\tau_1 = B$.
It is known that the form of the factorization theorem which holds for the thrust distribution breaks down for the broadening distribution. Thus, by varying $a$ between 0 and 1 it is possible to obtain a wealth of information on factorization~\cite{Hornig:2009kv} and  the final state, beyond what can be learned by looking at a single event shape in isolation. 

\vspace{3em}

This work was supported in part by the Director, Office of Science, Office of Nuclear Physics, of the U.S. Department of Energy under grant number DE-FG02-06ER41449.

\end{document}